\documentclass[11pt]{article}
\usepackage{slashed,cite}
\newsavebox{\uuunit} \sbox{\uuunit}
    {\setlength{\unitlength}{0.825em}
     \begin{picture}(0.6,0.7)
        \thinlines
        \put(0,0){\line(1,0){0.5}}
        \put(0.15,0){\line(0,1){0.7}}
        \put(0.35,0){\line(0,1){0.8}}
       \multiput(0.3,0.8)(-0.04,-0.02){12}{\rule{0.5pt}{0.5pt}}
     \end {picture}}
\newcommand {\unity}{\mathord{\!\usebox{\uuunit}}}
\def\rmi{{\rm i}}
\def\rmd{{\rm d}}
\newcommand{\ft}[2]{{\textstyle\frac{#1}{#2}}}
\newcommand{\hc}{{\rm h.c.}}
\def\Re{\mathop{\rm Re}\nolimits}
\def\Im{\mathop{\rm Im}\nolimits}

\csname @addtoreset\endcsname{equation}{section}
\begin{document}
\begin{titlepage}
\begin{flushright}
KUL-TF-04/14\\
hep-th/0405158
\end{flushright}
\vspace{.5cm}
\begin{center}
\baselineskip=16pt {\LARGE    Hypermultiplets and hypercomplex geometry\\
\vskip 0.2cm
from 6 to 3 dimensions
}\\
\vfill
{\Large Jan Rosseel and Antoine Van Proeyen $^\dagger$ 
  } \\
\vfill
{\small Instituut voor Theoretische Fysica, Katholieke Universiteit Leuven,\\
       Celestijnenlaan 200D, B-3001 Leuven, Belgium.
 }
\end{center}
\vfill
\begin{center}
{\bf Abstract}
\end{center}
{\small The formulation of hypermultiplets that has been developed for
5-dimensional matter multiplets is by dimensional reductions translated
into the appropriate spinor language for 6 and 4 dimensions. We also
treat the theories without actions that have the geometrical structure of
hypercomplex geometry. The latter is the generalization of hyper-K{\"a}hler
geometry that does not require a Hermitian metric and hence corresponds
to field equations without action. The translation tables of this paper
allow the direct application of superconformal tensor calculus for the
hypermultiplets using the available Weyl multiplets in 6 and 4
dimensions. Furthermore, the hypermultiplets in 3 dimensions that result
from reduction of vector multiplets in 4 dimensions are considered,
leading to a superconformal formulation of the \textbf{c}-map and an
expression for the main geometric quantities of the hyper-K{\"a}hler
manifolds in the image of this map.
 }\vspace{2mm} \vfill
 \hrule width 3.cm
 {\footnotesize \noindent
$^\dagger$  \{jan.rosseel, antoine.vanproeyen\}@fys.kuleuven.ac.be }
\end{titlepage}
\tableofcontents{}
\newpage
\section{Introduction}

It has been known for a long time that hypermultiplets in theories with 8
fermionic supercharges are related to quaternionic
geometry~\cite{Bagger:1983tt}, which has found various applications. In
the context of the construction of 5-dimensional supergravity-matter
couplings a new formulation for hypermultiplets has been
given~\cite{Bergshoeff:2002qk}. The starting point for this construction
is the realization of the supersymmetry algebra, rather than the
invariance of a proposed action. This new point of view leads to a more
manifest geometrical formulation of the hypermultiplets. Thereby the
ingredients of a hypercomplex structure are obtained first, before a
metric appears. The metric is introduced afterwards, when in a second
stage the action is considered, leading to hyper-K{\"a}hler manifolds. On the
physics side, this treatment has the advantage that theories whose field
equations are not necessarily derivable from an action principle are
included in this analysis. Moreover, there is a transparent way in which
supergravity couplings are obtained. A hypercomplex or hyper-K{\"a}hler
manifold can lead to a supergravity theory if it allows a conformal
structure, mathematically expressed as the presence of a closed
homothetic Killing vector. Such a vector allows a straightforward use of
the superconformal tensor calculus, leading to a supergravity theory as
has been illustrated in~\cite{Bergshoeff:2004kh}.

It turns out that the supergravity action and corresponding geometry are
completely fixed once the rigid one is determined, and this conformal
structure (closed homothetic Killing vector) is known. The supergravity
theory is obtained using superconformal tensor calculus (see for instance
\cite{Kaku:1977pa,Ferrara:1977ij,Kaku:1978ea,VanProeyen:1983wk}). This
construction goes by coupling the hypermultiplets (or other matter
multiplets) to a multiplet containing the gauge fields of the
superconformal symmetries, called the `Weyl multiplet'. After gauge
fixing conformal symmetries, one obtains the couplings of the matter
multiplet to Poincar{\'e} supergravity. If multiplets with gauge fields are
included, then also the couplings with gauged isometries are obtained
using the same steps. Especially for hypermultiplets, the latter
procedure will be clarified in~\cite{Bergshoeff:2003yy}.

It has been mentioned in~\cite{Bergshoeff:2002qk} that this development
is independent of the application to 5 dimensions, and could also be
applied to 6 and 4 dimensions. However, this generalization is not
straightforward due to the fact that spinors are described in $6$
dimensions as symplectic-Weyl spinors, in $5$ dimensions as symplectic
spinors and in $4$ dimensions as Majorana spinors. Therefore the
appearance of the geometric quantities differs (see for instance
\cite{deWit:1999fp}, where an analysis of rigid hypermultiplets with
action is performed in 4 dimensions, giving rise to hyper-K{\"a}hler
geometry). In order to be able to use the general analysis of
\cite{Bergshoeff:2002qk} in other dimensions, one needs a translation
table. Especially for applications of supersymmetry and supergravity in
$4$ dimensions such a translation is of practical use, and on the other
hand not straightforward. In this paper we will generalize the results
from \cite{Bergshoeff:2002qk} to $4$ and $6$ dimensions through
dimensional reduction of the transformation rules of the $5$-dimensional
theory. This will lead to connections between the different geometrical
quantities in the theories in 4, 5 and 6 dimensions. The advances made in
5-dimensional supersymmetry and supergravity are then immediately
applicable for 4 and 6 dimensions. Indeed, the superconformal tensor
calculus for 4
dimensions~\cite{deWit:1980ug,deRoo:1980mm,deWit:1981tn,deWit:1985px} and
6 dimensions~\cite{Bergshoeff:1986mz}, with especially the formulation of
the Weyl and vector multiplets that we need for this programme, has
already been known for a long time. Therefore, the formulation of the
basic elements of rigid hypermultiplets in this language is all that is
needed.

In the second part of this paper we will start from the four-dimensional
vector multiplet, whose geometry (known as special geometry) is fixed by
a holomorphic function of the complex scalars in the vector multiplet.
Reducing it to $3$ dimensions gives again a hypermultiplet. The
dimensional reduction from four to three dimensions in supergravity gives
rise to the \textbf{c}-map between special K{\"a}hler geometry and
quaternionic-K{\"a}hler geometry. This map was first introduced
in~\cite{Cecotti:1989qn}. While it is obtained in supergravity by
dimensional reduction from 4 to 3 dimensions, it is obtained in string
theory by a T-duality between type IIA and type IIB strings before
dimensionally reducing both string theories from 10 to 4 dimensions. This
procedure exchanges vector multiplets with hypermultiplets and thus
provides a mapping between special K{\"a}hler and quaternionic-K{\"a}hler
geometry. This \textbf{c}-map leads to the notion of `{\it special
quaternionic-K{\"a}hler manifolds}', which are those manifolds appearing in
the image of the \textbf{c}-map. They are a subclass of the
quaternionic-K{\"a}hler manifolds. In rigid supersymmetry a similar map has
been considered in~\cite{DeJaegher:1998ka} by dimensional reduction from
4 to 3 dimensions. The resulting subclass of hyper-K{\"a}hler manifolds in
the image of this map has been called `special hyper-K{\"a}hler'
manifolds\footnote{Observe that the \textbf{c}-map for rigid K{\"a}hler
geometry does not lead directly to the \textbf{c}-map for local K{\"a}hler
geometry. The latter increases the dimension from $n$ complex to $n+1$
quaternionic. The extra quaternion originates from the dimensional
reduction of pure supergravity in 4 dimensions. The map that we consider
here does therefore \textit{not} lead to a hyper-K{\"a}hler manifold that can
be mapped to a quaternionic-K{\"a}hler manifold by using superconformal
tensor calculus.}. We will reconsider this map using the geometric
formulation as first developed in 5 dimensions and using the same
generalization to other dimensions as in the first part of the paper.

In section \ref{sechyper} we will start from the 5-dimensional
hypermultiplet. After reducing it to 4 dimensions we obtain translations
of the geometric quantities defining the hypercomplex geometry. Then we
will do the same analysis again for the 6-dimensional hypermultiplet.
Equations of motion are derived through closure of the supersymmetry
algebra.

Section \ref{secvechyper} starts with a summary of the dimensional
reduction of the rigid supersymmetry transformations of the 4-dimensional
vector multiplet to 3 dimensions~\cite{DeJaegher:1998ka}, leading to the
rigid \textbf{c}-map discussed above. We will use the obtained
translation formulae to lift the resulting hypermultiplet up to 5
dimensions. This facilitates the identification of the main geometrical
quantities, due to the more transparent notation in 5 dimensions. We
obtain in this way general formulae for the curvature tensors of special
hyper-K{\"a}hler manifolds.

\section{Hypermultiplets in $d=5,4,6$} \label{sechyper}

In this section, we review hypermultiplets in $4, 5$ and $6$ dimensions,
without assuming the existence of an action. We start by considering the
transformations of the hypermultiplet in $d=5$ and $4$ under rigid
supersymmetry. Then we reduce the five-dimensional theory to $4$
dimensions. This allows us to relate the geometrical quantities that
arise in both dimensions. Next, we lift the five-dimensional theory up to
$d=6$. In this way, we are able to show that each time, the scalars
describe a hypercomplex manifold, which appears in different guises,
depending on the dimension considered. We also derive equations of motion
for the fermions through closure of the supersymmetry algebra.

\subsection{Hypermultiplets in $d=5$}
\label{ss:hyperd5}

Here we review the description of hypermultiplets in five dimensions. An
interesting discussion of this can be found in \cite{Bergshoeff:2002qk}.
For calculations with spinors we will mainly use the conventions of
\cite{VanProeyen:1999ni}.

A system of $r$ hypermultiplets consists of $4r$ real scalars $q^X(x)$,
$X = 1, \cdots, 4r$ and $2r$ spinors $\zeta^A(x)$, $A = 1, \cdots, 2r$.
In five dimensions, the spinors are subject to symplectic Majorana
reality conditions. One introduces matrices $\rho_A{}^B$ and $E_i{}^j$,
obeying
\begin{equation}
\rho \rho^* = -\unity _{2r}, \qquad EE^* = -\unity _2.
\label{rho2isunity}
\end{equation}
Using these, symplectic Majorana conditions for the fermions and
supersymmetry parameters can be defined as
\begin{equation} \label{symmajzeta}
(\zeta^A)^* = \alpha \mathcal{C}_5 \gamma_0 \zeta^B \rho_B{}^A, \qquad
(\epsilon^i)^* = \alpha \mathcal{C}_5 \gamma_0 \epsilon^j E_j{}^i,
\end{equation}
where $\mathcal{C}_5$ is the charge conjugation matrix in 5 dimensions
and $\alpha$ is an irrelevant number of modulus $1$. We will always use
the basis where $E_i{}^j = \varepsilon_{ij}$. Indices are raised and
lowered using the north-west-south-east convention.

The supersymmetry transformations take the general form
\begin{eqnarray} \label{susyD5}
\delta(\epsilon)q^X & = & -\rmi \bar{\epsilon}^i \zeta^A f^X_{iA},
\nonumber \\
\delta(\epsilon)\zeta^A & = & \frac{1}{2} \rmi \gamma^\mu \partial_\mu
q^X f^{iA}_X \epsilon_i - \zeta^B \omega_{XB}{}^A (\delta(\epsilon)q^X).
\end{eqnarray}
Here, $f_{iA}^X(q)$, $f^{iA}_X(q)$ and $\omega_{XB}{}^A(q)$ are arbitrary
functions of the scalars. Due to reality of the scalars and the
symplectic Majorana conditions for
the fermions, they satisfy the following reality conditions
\begin{equation} \label{realfomega}
\big(f^{iA}_X\big)^* = f^{jB}_X E_j{}^i \rho_B{}^A, \qquad
\big(\omega_{XA}{}^B\big)^* = (\rho^{-1} \omega_X \rho)_A{}^B.
\end{equation}
In order that the commutator of $2$ supersymmetry transformations
gives a translation, one has to impose
\begin{eqnarray} \label{commqrestr}
f^{iA}_Y f^X_{iA} & = & \delta_Y^X,\qquad f^{iA}_X f^X_{jB}  =
 \delta_j^i \delta^A_B, \nonumber \\
\mathcal{D}_Y f^X_{iB} & \equiv & \partial_Y f^X_{iB} - \omega_{YB}{}^A
f^X_{iA} + \Gamma_{ZY}{}^X f^Z_{iB} = 0,
\end{eqnarray}
where $\Gamma$ is an object symmetric in its lower indices. Note that
$f^X_{iA}$ and $f^{iA}_X$ are each others inverse and that they are
covariantly constant with connections $\Gamma$ and $\omega$. These are
the only conditions on the target space that follow from imposing closure
of the supersymmetry transformations. There are no further geometrical
constraints from the commutator of 2 supersymmetries on the fermions. In
this respect, this commutator will rather define equations of motion for
the hypermultiplet.

\paragraph{Geometry.} The geometry of the target space is a
hypercomplex manifold. In fact it is a weakened form of hyper-K{\"a}hler
geometry, where no Hermitian covariantly constant metric is defined. The
basic object for defining these manifolds is a triplet of complex
structures, the hypercomplex structure
\begin{equation} \label{complstr}
\vec J_X{}^Y \equiv -\rmi f^{iA}_X \vec \sigma_i{}^j f^Y_{jA}.
\end{equation}
These structures are covariantly constant and satisfy the quaternion
algebra. The statement for arbitrary 3-vectors $\vec \alpha $ and $\vec
\beta $,
\begin{equation} \label{quatalg}
\vec \alpha \cdot \vec J\,\vec \beta \cdot \vec  J = -\unity _{4r} \vec
\alpha \cdot \vec \beta + \left( \vec \alpha \times \vec \beta\right)
\cdot \vec J.
\end{equation}
We adopt the following definitions for the curvature tensors of the $\Gamma$ and
$\omega$ connections:
\begin{eqnarray} \label{defcurvatures}
R_{XYZ}{}^W & \equiv & 2 \partial_{[X}\Gamma_{Y]Z}{}^W + 2 \Gamma_{V[X}{}^W
\Gamma_{Y]Z}{}^V, \\
\mathcal{R}_{XYB}{}^A & \equiv & 2 \partial_{[X}\omega_{Y]B}{}^A + 2
\omega_{[X|C|}{}^A \omega_{Y]B}{}^C.
\end{eqnarray}
The integrability conditions on the vielbeins $f^{iA}_X$ relate the
curvature tensors $R_{XYZ}{}^W$ and $\mathcal{R}_{XYB}{}^A$:
\begin{equation} \label{RGROm}
R_{XYZ}{}^W = f_{iA}^W f^{iB}_Z \mathcal{R}_{XYB}{}^A, \qquad \delta_j^i
\mathcal{R}_{XYB}{}^A = f_W^{iA} f^Z_{jB} R_{XYZ}{}^W.
\end{equation}
Using cyclicity properties of the Riemann tensor, one can also obtain
\begin{eqnarray} \label{defW}
f^X_{Ci} f^Y_{jD} \mathcal{R}_{XYB}{}^A & = & -\ft{1}{2} \varepsilon_{ij}
W_{CDB}{}^A, \nonumber \\
W_{CDB}{}^A & \equiv & f_C^{iX} f^Y_{iD} \mathcal{R}_{XYB}{}^A =
\ft{1}{2}f_C^{iX} f^Y_{iD} f_{jB}^Z f^{Aj}_W R_{XYZ}{}^W.
\end{eqnarray}
This tensor $W$ is symmetric in its $3$ lower indices.

\paragraph{Dynamics.}
After calculating the commutator of 2 supersymmetries on the fermions,
one notes that the algebra does not close. Demanding that non-closure
terms vanish leads to equations of motion for the fermions:
\begin{equation} \label{Gamma}
\Gamma^A = \gamma^a \mathcal{D}_a \zeta^A + \ft{1}{2} W_{CDB}{}^A \zeta^B
\bar{\zeta}^D \zeta^C = 0,
\end{equation}
where the covariant derivative is given by
\begin{equation} \label{coderzeta}
\mathcal{D}_\mu \zeta^A \equiv \partial_\mu \zeta^A + (\partial_\mu q^X)
\zeta^B \omega_{XB}{}^A.
\end{equation}
So one sees that imposing the supersymmetry algebra on the scalars leads
to defining the hypercomplex geometry, while closure on the fermions
leads to equations of motion for these fermions.

\paragraph{Projection to supergravity.} A closed homothetic Killing
vector is a vector $k^X$ satisfying
\begin{equation}
  \mathcal{D}_Y k^X=\frac{d-2}2\delta _Y{}^X=\frac32\delta _Y{}^X.
 \label{closedhomoth}
\end{equation}
This induces a conformal symmetry, where e.g. the dilatations with
parameter $\lambda _D$ act on the hypermultiplet scalar as
\begin{equation}
  \delta _D(\lambda _D) q^X=\lambda _D \left( x^\mu \partial _\mu q^X +
  k^X\right).
 \label{DonqX}
\end{equation}
The vector $k^X$ and the 3 vectors $k^Y\vec J_Y{}^X$ define 4 scalars
that form a quaternion. After gauge-fixing the superconformal symmetry
using the Weyl multiplet, this quaternion is gauge-fixed, together with
its fermionic partner, such that a supergravity theory coupled to
hypermultiplets remains. Clearly the hypermultiplet sector has one less
quaternionic dimension in the supergravity theory than in the
corresponding rigid supersymmetry theory that was its starting point.
This is analysed in detail in~\cite{Bergshoeff:2003yy}.

\subsection{Hypermultiplets in $d=4$}
\label{ss:hyperd4}

Superconformal hypermultiplets in four dimensions were discussed in
\cite{deWit:1999fp}. We will review the supersymmetry transformations and
the resulting hypercomplex geometry and equations of motion.

The main difference with the $5$-dimensional case lies in the reality
conditions that one imposes on the fermions. Whereas in five dimensions
one is obliged to impose symplectic Majorana conditions, in four
dimensions one has Majorana spinors. Moreover, one can write the Majorana
spinors as chiral spinors. So a system of $r$ hypermultiplets will now
consist of $4r$ real scalars $q^X$, and $2r$ spinors
$\zeta^{\bar{\alpha}}$ with positive chirality and $2r$ spinors
$\zeta^\alpha$ with negative chirality. By complex conjugation, the
indices $\alpha$ and $\bar{\alpha}$ are interchanged, so we have indeed a
system of $2r$ Majorana spinors. The chirality of the $4$-dimensional
supersymmetry parameters is indicated by the position of the
SU($2$)-index
\begin{equation} \label{convchir4}
\tilde{\epsilon}^i = \gamma_5 \tilde{\epsilon}^i, \qquad
\tilde{\epsilon}_i = - \gamma_5 \tilde{\epsilon}_i.
\end{equation}

The general form of the supersymmetry transformations is then
\begin{eqnarray} \label{susytransf4}
\delta(\tilde{\epsilon})q^X & = & \gamma^X_{i\bar{\alpha}}
\bar{\tilde{\epsilon}}^i\zeta^{\bar{\alpha}} +
\bar{\gamma}^{Xi}_{\alpha}\bar{\tilde{\epsilon}}_i \zeta^{\alpha},
\nonumber \\
\delta(\tilde{\epsilon}) \zeta^\alpha & = & V^\alpha_{Xi} \gamma^\mu
\partial_\mu q^X \tilde{\epsilon}^i - \delta(\tilde{\epsilon})q^X
\Gamma_X{}^\alpha{}_\beta \zeta^\beta, \nonumber \\
\delta(\tilde{\epsilon}) \zeta^{\bar{\alpha}} & = & \bar{V}^{i
\bar{\alpha}}_{X} \gamma^\mu
\partial_\mu q^X \tilde{\epsilon}_i - \delta(\tilde{\epsilon})q^X
\bar{\Gamma}_X{}^{\bar{\alpha}}{}_{\bar{\beta}} \zeta^{\bar{\beta}}.
\end{eqnarray}
Again the coefficients $\gamma^X_{i\bar{\alpha}}$, $V^\alpha_{Xi}$ and
$\Gamma_X{}^\alpha{}_\beta$ are functions of the scalars $q^X$. Their
complex conjugates are
\begin{equation}
  \bar{\gamma}^{Xi}_\alpha=\left(\gamma^X_{i\bar{\alpha}}\right) ^*, \qquad
\bar{V}^{i \bar{\alpha}}_X=\left(V^\alpha_{Xi}\right)^*, \qquad
\bar{\Gamma}_X{}^{\bar{\alpha}}{}_{\bar{\beta}}=\left(\Gamma_X{}^\alpha{}_\beta\right)
^*.
 \label{ccvielbeins}
\end{equation}
Note that complex conjugation raises or lowers here the indices $i,j$,
while indices $\alpha $ are replaced by $\bar \alpha $. Demanding that
the commutator of $2$ supersymmetries on the scalars gives a translation,
leads to the restrictions
\begin{eqnarray} \label{cliffcondeninvertvielb4}
& & \gamma^X_{i \bar{\alpha}} \bar{V}^{j \bar{\alpha}}_Y +
\bar{\gamma}^{X j}_{\alpha} V^\alpha_{Y i} = \delta^j_i \delta^X_Y,
\nonumber \\
& & \bar{V}^{i \bar{\alpha}}_X \gamma^X_{j \bar{\beta}} = \delta^i_j
\delta^{\bar{\alpha}}_{\bar{\beta}}.
\end{eqnarray}
The first condition is also known as the Clifford condition. The second
condition expresses the invertibility of the vielbeins. Demanding closure
of the superalgebra leads to the following relations:
\begin{eqnarray} \label{covconstviel4}
\partial_Y \gamma^X_{i \bar{\alpha}} + \Gamma_{YZ}{}^X \gamma^Z_{i
\bar{\alpha}} - \gamma^X_{i \bar{\beta}}
\Gamma_Y{}^{\bar{\beta}}{}_{\bar{\alpha}} & = & 0, \nonumber \\
\partial_Y \bar{\gamma}^{Xi}_\alpha + \Gamma_{YZ}{}^X
\bar{\gamma}^{Zi}_\alpha - \bar{\gamma}^{Xi}_\beta
\Gamma_Y{}^\beta{}_\alpha & = & 0.
\end{eqnarray}
Again the $\Gamma_{XY}{}^Z$ are symmetric in the lower indices ($XY$). So
we see that the vielbeins $\gamma$ (or $\bar{\gamma}$) are covariantly
constant with respect to connections $\Gamma_X{}^\alpha{}_\beta$ (or
$\bar{\Gamma}_X{}^{\bar{\alpha}}{}_{\bar{\beta}}$) and $\Gamma_{XY}{}^Z$.
Again these are the only conditions implied by supersymmetry on the
target space. They imply that the manifold is hypercomplex. Closure of
the supersymmetry algebra on the fermions leads to equations of motion.

One can define a matrix $\rho$:
\begin{equation} \label{defrho4}
\rho^\alpha{}_{\bar{\beta}} \equiv \ft{1}{2} \varepsilon^{ij}
V_{Xi}^\alpha \gamma^X_{j{\bar{\beta}}},
\end{equation}
that is subject to the following conditions:
\begin{equation} \label{eigrho4}
\rho^\alpha{}_{\bar{\beta}} \rho^{\bar{\beta}}{}_\gamma =
-\delta_\gamma^\alpha, \qquad \bar{\gamma}^{Xi}_\alpha
\rho^\alpha{}_{\bar{\beta}} \varepsilon_{ij} =
\gamma^X_{j\bar{\beta}},\qquad \varepsilon_{ij}
\rho^\alpha{}_{\bar{\beta}} \bar{V}_X^{j\bar{\beta}} = V^\alpha_{Xi}.
\end{equation}
The last two equations are reality conditions for the $\gamma$ and $V$
coefficients. In this respect, this matrix $\rho $ is similar to that
used in~(\ref{symmajzeta}), and satisfies the same
relation~(\ref{rho2isunity}). However, in 4 dimensions the complex
conjugation of the vielbein coefficients is defined
in~(\ref{ccvielbeins}), and $\rho $ is defined by~(\ref{defrho4}) while
in 5 dimensions the vielbeins and their complex conjugates are in the
same tensors $f^{iA}_X$ related by $\rho $ as
in~(\ref{realfomega}).\footnote{This difference is relevant when
comparing the calculation of commutators on scalars between 4 and 5
dimensions. Note that the result~(\ref{commqrestr}) is already obtained
from the commutator on the scalars in 5 dimensions, while in 4 dimensions
we need the commutators on scalars and fermions to arrive
at~(\ref{cliffcondeninvertvielb4}). The difference is that the reality
relation~(\ref{realfomega}) follows from the properties of symplectic
Majorana spinors in 5 dimensions. The analogous relation in 4 dimensions
is only derived after the further steps explained above.}

The discussion in \cite{deWit:1999fp} implied the existence of an action
and thus a metric for the target space. The resulting geometry was that
of a hyper-K{\"a}hler manifold. Here we will no longer demand the existence
of an action. We expect to find again the hypercomplex manifold discussed
in the previous section. Again, we expect to find a triplet of complex
structures $\vec J$ that satisfies the quaternion algebra
(\ref{quatalg}). One can also calculate the necessary curvature tensors.
The integrability condition on the vielbeins now leads to
\begin{eqnarray} \label{krommingsrelaties4}
R_{XYV}{}^Z \bar{\gamma}^{Vi}_\beta V^\gamma_{jZ} & = & \delta_j^i
\mathcal{R}_{XY}{}^\gamma{}_\beta, \nonumber \\
R_{XYV}{}^Z \gamma^V_{i\bar{\beta}} \bar{V}^{j\bar{\gamma}}_Z & = &
\delta_i^j \mathcal{R}_{XY}{}^{\bar{\gamma}}{}_{\bar{\beta}},
\end{eqnarray}
where the curvature tensors are defined in a way similar to that in five
dimensions.

The curvature tensors are dependent on a conformal tensor $W$, which we
also introduced in the $5$-dimensional case. It determines both
$\mathcal{R}_{XY}{}^\alpha{}_\beta$ and $R_{XYZ}{}^W$. It can be defined
by
\begin{equation} \label{defW4}
W_{\gamma \delta}{}^\alpha{}_\beta \equiv \ft{1}{4} \varepsilon_{ij}
\bar{\gamma}^{Xj}_\gamma \bar{\gamma}^{Yi}_\delta
\mathcal{R}_{XY}{}^\alpha{}_\beta = \ft{1}{8} \varepsilon_{ij}
\bar{\gamma}^{Xj}_\gamma \bar{\gamma}^{Yi}_\delta \bar{\gamma}^{Vk}_\beta
V^\alpha_{kZ} R_{XYV}{}^Z.
\end{equation}
(We take normalizations such that it is the dimensional reduction of the
definition that we have in 5 dimensions, see below.) The tensor is
symmetric in the 3 lower indices, as follows from cyclicity properties of
the Riemann tensor. One can also define this tensor with
$\bar{\alpha}$-indices instead of $\alpha$-indices by multiplication with
$\rho^\kappa{}_{\bar{\alpha}}$. For instance
\begin{equation}
W_{\bar{\alpha} \beta}{}^\gamma{}_\delta = W_{\kappa
\beta}{}^\gamma{}_\delta \rho^\kappa{}_{\bar{\alpha}} = -\ft{1}{4}
\gamma^X_{i \bar{\alpha}} \bar{\gamma}^{Yi}_\beta
\mathcal{R}_{XY}{}^\gamma{}_\delta. \label{Wbarnobar}
\end{equation}

As in the five-dimensional case, calculating the commutator of $2$
supersymmetry transformations on the fermions leads to non-closure terms,
corresponding to dynamical equations for these fermions. One obtains
\begin{equation}
\gamma^\mu \mathcal{D}_\mu \zeta^\alpha + \ft{1}{2} W_{\bar{\gamma}
\delta}{}^\alpha{}_\beta \zeta^{\bar{\gamma}} \bar{\zeta}^\delta
\zeta^\beta = 0.
\end{equation}
The covariant derivative is given by
\begin{equation}
\mathcal{D}_\mu \zeta^\alpha = \partial_\mu \zeta^\alpha + (\partial_\mu
q^X) \Gamma_{X}{}^{\alpha}{}_{\beta} \zeta^\beta.
\end{equation}

\subsection{Dimensional reduction from $5$ to $4$ dimensions}
In this section, we perform the process of dimensional reduction in order
to obtain the $4$-dimensional theory from the $5$-dimensional one. We
first give a brief sketch of how this procedure works. Then we apply this
to obtain relations between the geometrical quantities appearing in both
dimensions. Finally, we show how these relations lead to translations of
the constraints defining the hypercomplex geometry in the different
dimensions. This will allow us to conclude that indeed the same
geometrical relations appear, but in different guises.

\subsubsection{Some notes on the reduction}
In order to perform the reduction, we will suppose that the fields are
independent of the fifth spacetime coordinate. We also have a set of five
$\gamma$-matrices. In four dimensions, we will use the first four of
these matrices to form the four-dimensional Clifford algebra. The fifth
$\gamma$-matrix will be used to form projection operators $P_L =
\frac{1}{2}(1+\gamma_5)$ and $P_R = \frac{1}{2}(1-\gamma_5)$, allowing us
to split $4$-dimensional spinors into chiral spinors. In $5$ dimensions
the spinors $\zeta^A(x)$ are subject to symplectic Majorana conditions,
while in $4$ dimensions we assume Majorana reality conditions for the
fermion fields. The reduction can be obtained by taking the following
identification between the $5$- and $4$-dimensional spinors:
\begin{eqnarray} \label{idzeta}
\sqrt{2}\zeta^{\bar{\alpha}} & = & \ft{1}{2} (1+\gamma_5) \zeta^A,
\nonumber
\\
\sqrt{2}\zeta^{\alpha} & = & \ft{1}{2} (1-\gamma_5) \zeta^B \rho_B{}^A.
\end{eqnarray}
Note that we use here a special notation where the value of $A$ is the
same as that of $\alpha$ or in other cases of $\bar{\alpha}$ (complex
conjugate). We still keep the different index notation, as that will make
it clear whether the object is a quantity that appears in $5$ dimensions
or in $4$ dimensions. Note that the four-dimensional charge conjugation
matrix is given by $\mathcal{C}_4 = \mathcal{C}_5 \gamma_5$. A similar
identification as in (\ref{idzeta}) applies to the supersymmetry
parameters
\begin{eqnarray} \label{ideps}
\sqrt{2}\tilde{\epsilon}^i & = & \ft{1}{2} (1+\gamma_5) \epsilon^i,
\nonumber \\
\sqrt{2}\tilde{\epsilon}_i & = & \ft{1}{2} (1-\gamma_5) \epsilon^j
E_j{}^i.
\end{eqnarray}

\subsubsection{Connections between $5$- and $4$-dimensional
quantities}

The scalars $q^X$ of the 4 and 5-dimensional hypermultiplets are
trivially identified, and reducing the supersymmetry transformation laws
according to the above rules, we can relate the geometrical quantities
appearing in the 4 and 5-dimensional theories. We obtain the following
translation formulae between the vielbeins appearing in the supersymmetry
transformation rules in the different dimensions considered:
\begin{equation} \label{vielbeins54}
f^X_{iA}  =  \frac{\rmi}{2}\gamma^X_{i\bar{\alpha}} =
\frac{\rmi}{2}\rho^\beta{}_{\bar{\alpha}}
\bar{\gamma}^{Xj}_{\beta}\varepsilon_{ji}.
\end{equation}
For the inverse vielbeins we get
\begin{equation} \label{invvielbeins54}
f^{iA}_X  =  -2\rmi \bar{V}^{i\bar{\alpha}}_X = 2\rmi \varepsilon^{ki}
\rho^{\bar{\alpha}}{}_{\beta} V^\beta_{Xk}.
\end{equation}
We can also relate the spin connections in the 4 and 5-dimensional
formalism
\begin{equation} \label{spinconn45}
\bar{\Gamma}_X{}^{\bar{\alpha}}{}_{\bar{\beta}}  = \omega_{XB}{}^A,
\qquad \Gamma_X{}^\delta{}_\gamma  =  (\rho^{-1} \omega_X \rho)_C{}^D.
\end{equation}
The reality conditions of the different quantities of the
five-dimensional theory are consistent with those of the quantities of
the 4-dimensional formalism. Thus, we have obtained the necessary
relations between the fundamental quantities defining the hypermultiplet.

\subsubsection{Reduction of geometrical quantities}

In this section, we will apply the previously derived formulae to relate
the geometrical constraints and quantities in the different languages. By
using the identifications (\ref{vielbeins54}) and (\ref{invvielbeins54})
in the equation $f^{iA}_X f^X_{jB} = \delta^i_j \delta^A_B$, one obtains
the second condition in (\ref{cliffcondeninvertvielb4}). By using the
same vielbein identifications and  the first relation of
(\ref{commqrestr}) one can get the Clifford condition of the
four-dimensional theory. Indeed, first for $i = j$, using the obtained
identifications, we get
\begin{equation}
\gamma^X_{i\bar{\alpha}} \bar{V}^{i\bar{\alpha}}_Y +
\bar{\gamma}^{Xi}_\alpha V^\alpha_{Yi} = f^X_{iA} f^{iA}_Y + f^X_{lB}
f^{lB}_Y = 2 \delta^X_Y.
\end{equation}
Considering the same condition with $i \neq j$ gives
\begin{equation}
\gamma^X_{i\bar{\alpha}} \bar{V}^{j\bar{\alpha}}_Y +
\bar{\gamma}^{Xj}_\alpha V^\alpha_{Yi} = f^{jA}_Y f^X_{iA} - f^{jB}_Y
f^X_{iB},
\end{equation}
as expected. It is also easy to see that the covariant constancy of the
vielbeins in the four-dimensional formalism is just the translation of
the covariant constancy of the vielbeins in the five-dimensional theory.

{}From the identifications (\ref{spinconn45}) for the spin connections,
it follows directly that the definition of
$\mathcal{R}_{XY}{}^{\bar{\alpha}}{}_{\bar{\beta}}$ is the translation of
the definition of $\mathcal{R}_{XYB}{}^A$. Using the vielbein
identifications, one also obtains (\ref{krommingsrelaties4}) by
translating (\ref{RGROm}). One can also show that the curvature tensor
$\mathcal{R}_{XY}{}^\alpha{}_\beta$ corresponds to the spin curvature of
the 5-dimensional formalism in the following way:
\begin{equation}
\mathcal{R}_{XY}{}^\beta{}_\alpha = (\rho^{-1} \mathcal{R}_{XY}
\rho)_A{}^B.
\end{equation}
Note that in five dimensions
\begin{equation}
\big(\mathcal{R}_{XYA}{}^B\big)^* = (\rho^{-1} \mathcal{R}_{XY}
\rho)_A{}^B.
\end{equation}
Hence, we have
\begin{equation}
 \mathcal{R}_{XYA}{}^B =\mathcal{R}_{XY}{}^{\bar \beta}{}_{\bar
 \alpha}.
 \label{RGld5d4}
\end{equation}
As mentioned already, the matrix $\rho $ defined in~(\ref{defrho4}) is
related to the $\rho$-matrix, used in five dimensions to define the
symplectic Majorana conditions. It turns out that
\begin{equation}
\rho_B{}^A  =  \rho^{\alpha}{}_{\bar{\beta}},\qquad
\big(\rho_B{}^A\big)^* =
 \rho^{\bar{\alpha}}{}_\beta.
\end{equation}
In this way, one can obtain the conditions (\ref{eigrho4}) by translating
the reality conditions for the vielbeins in 5 dimensions
(\ref{realfomega}).

As mentioned, the scalars $q^X$ of the 4- and 5-dimensional
hypermultiplets are identified. Bosonic quantities as the complex
structures can therefore be straightforwardly identified too. The
expressions in terms of frame-dependent quantities as in (\ref{complstr})
are, however, different. The translation of the latter into the
$4$-dimensional formalism gives
\begin{equation}
\vec J_X{}^Y = -\rmi \bar{V}_X^{j\bar{\alpha}} \vec \sigma_j{}^i
\gamma^Y_{i\bar{\alpha}}.
\end{equation}

The conformal tensor $W$ as defined in~(\ref{defW4}) is the translation
of the five-dimensional $W$-tensor
\begin{equation}
  W_{ABC}{}^D=W_{\bar \alpha \bar \beta }{}^{\bar \delta }{}_{\bar \gamma
  }.
 \label{Wd5d4}
\end{equation}
The translation of the holomorphic $W$-tensor (with $\alpha$-indices) can
be obtained by multiplication with $\rho^\kappa{}_{\bar{\alpha}}$,
see~(\ref{Wbarnobar}).

It is useful to see that this translation suggests a simplification in
the formulation of hypermultiplets in 4 dimensions. Indeed, it turns out
that we can restrict ourselves to the quantities with only $\bar \alpha$
indices, and no $\alpha $-indices to describe the full geometry. This
$\bar \alpha$ index is identified with the $A$ index of the formulation
in 5 dimensions. The basic relation is only the second line of
(\ref{cliffcondeninvertvielb4}). The objects with $\alpha $ indices are
defined as the complex conjugates, see (\ref{ccvielbeins}), and satisfy
as in 5 dimensions the reality constraint (\ref{eigrho4}). This is e.g.
sufficient to derive the first line of (\ref{cliffcondeninvertvielb4}).
The objects with $\alpha $-indices are useful for writing fermions with
negative chirality, but as the latter are related to the
positive-chirality components by the Majorana condition, they are not
independent.

\subsection{Hypermultiplets in $6$ dimensions}

In this section, we will obtain a theory of hypermultiplets in $d=6$,
starting from the $5$-dimensional theory. Again, special attention goes
to the geometry that arises in these theories. We start by giving some
comments about the dimensional reduction. Next, we argue that the
constraints implied by supersymmetry on the target space are the same as
in 5 dimensions. Finally, equations of motion are derived, without
assuming the existence of an action.

\subsubsection{Comments on uplifting from $5$ to $6$ dimensions}
In $5$ dimensions, a Clifford algebra of $\gamma$-matrices can be
represented by $4 \times 4$-matrices. In six dimensions, one now has $6$
$\Gamma$-matrices, which are $8 \times 8$-matrices. By using the
$5$-dimensional $\gamma$-matrices, one can construct the following
representation of the $6$-dimensional Clifford algebra.
\begin{eqnarray} \label{reprGamma}
\Gamma_0 & = & \left( \begin{array}{cc} \gamma_0 & 0 \\
0 & \gamma_0 \end{array} \right), \qquad \Gamma_1 = \left( \begin{array}{cc} \gamma_1 & 0 \\
0 & \gamma_1 \end{array} \right), \qquad \Gamma_2 = \left( \begin{array}{cc} \gamma_2 & 0 \\
0 & \gamma_2 \end{array} \right), \nonumber\\
 \Gamma_3 & = & \left( \begin{array}{cc} \gamma_3 & 0 \\
0 & \gamma_3 \end{array} \right), \qquad \Gamma_5 = \left( \begin{array}{cc} 0 & \gamma_5 \\
\gamma_5 & 0 \end{array} \right), \qquad \Gamma_6 = \left( \begin{array}{cc} 0 &
-\rmi \gamma_5 \\
\rmi \gamma_5 & 0 \end{array} \right).
\end{eqnarray}
One can also define a matrix $\Gamma_*$:
\begin{equation}
\Gamma_* = \Gamma_0\Gamma_1\Gamma_2\Gamma_3\Gamma_5\Gamma_6 =
\left( \begin{array}{cc} \gamma_5 & 0 \\ 0 & -\gamma_5
\end{array} \right).
\end{equation}
This matrix anticommutes with the other $\Gamma$-matrices.\\
One can also define a charge conjugation matrix $\mathcal{C}_6$ in
terms of $\mathcal{C}_5$:
\begin{equation}
\mathcal{C}_6 = \left( \begin{array}{cc} 0 & \mathcal{C}_5 \\
\mathcal{C}_5 & 0 \end{array} \right).
\end{equation}
In six dimensions, spinors also obey symplectic Majorana reality
conditions. Therefore, identifications between the $5$- and
$6$-dimensional spinors are simpler than in the previous case\footnote{We
use again $\tilde{\epsilon}$, $\tilde{\zeta}$ to denote six-dimensional
quantities, like in the four-dimensional discussion. As we will not
directly link 6- and 4-dimensional quantities, we hope that this will not
lead to confusion.}. We take a basis in which $\gamma_5$ is diagonal
[$\gamma _5=\mathop{\rm diag}\nolimits(1,1,-1,-1)$]. The spinors are
mapped as follows:
\begin{eqnarray}
\zeta ^A=\left(
\begin{array}{c} \zeta^A_1 \\ \zeta^A_2  \\
\zeta^A_3 \\ \zeta^A_4 \end{array} \right)& \rightarrow & \tilde{\zeta}^A
= \frac{1}{2} (1 + \Gamma_*) \tilde{\zeta}^A = \left(
\begin{array}{c} \zeta^A_1 \\ \zeta^A_2 \\ 0 \\ 0 \\ 0 \\ 0 \\
\zeta^A_3 \\ \zeta^A_4 \end{array} \right), \nonumber\\
\epsilon ^i=\left( \begin{array}{c} \epsilon^i_1 \\ \epsilon^i_2\\ \epsilon^i_3 \\
\epsilon^i_4
 \end{array} \right)& \rightarrow &
 \tilde{\epsilon}^i = \frac{1}{2} (1 - \Gamma_*) \,\tilde{\epsilon}^i\,
= \left( \begin{array}{c} 0 \\ 0 \\ \epsilon^i_3 \\ \epsilon^i_4
\\ \epsilon^i_1 \\ \epsilon^i_2 \\ 0 \\ 0 \end{array} \right),
\end{eqnarray}
where the tilde denotes the six-dimensional spinors, which are chiral (or antichiral).

The six-dimensional reality conditions are consistent with the
five-dimensional ones, due to the choice of the antisymmetric charge
conjugation matrix. One can then take the following supersymmetry
transformation rules for the six-dimensional hypermultiplet:
\begin{eqnarray} \label{transfin6}
\delta(\tilde{\epsilon}) q^X & = & -\rmi \bar{\tilde{\epsilon}}^i
\tilde{\zeta}^A
f^X_{iA}, \nonumber \\
\delta(\tilde{\epsilon}) \tilde{\zeta}^A & = & \ft{1}{2} \rmi \Gamma^a
\partial_a q^X f^{iA}_X \tilde{\epsilon}^j \varepsilon_{ji} -
\tilde{\zeta}^B \omega_{XB}{}^A (\delta(\tilde{\epsilon}) q^X).
\end{eqnarray}
One immediately sees that one can take the vielbeins to be the same
functions of the scalars as in $d=5$. The identification between the
vielbeins and spin connections in the $5$- and $6$-dimensional formalism
is thus very simple.

\subsubsection{Geometrical aspects}

The fact that the identifications between vielbeins and spin connections
are trivial  shows that the geometrical relations of the $5$-dimensional
formulation should also hold in the $6$-dimensional formulation. Indeed,
one can calculate the commutator of $2$ supersymmetry transformations on
the scalars and make sure that it gives a translation and closes. The
conclusions are the same as in the $5$-dimensional case. The vielbeins
obey the constraints (\ref{commqrestr}). As in the five-dimensional case,
these are the only restrictions on the target space of the scalars,
coming from imposing the supersymmetry algebra. One can take the same
definition (\ref{complstr}) for the complex structures. Integrability
conditions on the vielbeins and cyclicity properties of the Riemann
tensor lead to the same curvature relations (\ref{RGROm}) and conclusions
about the tensor $W$ (\ref{defW}). Again one obtains a hypercomplex
manifold.

\subsubsection{Dynamical aspects}

As already mentioned, the commutator of $2$ supersymmetries on the
fermions does not lead to new geometrical restrictions. Non-closure terms
will lead to equations of motion in the geometry. The calculation is
simpler than in the $5$-dimensional case, due to the chirality of the
spinors. We get
\begin{equation} \label{rescommferm6}
[\delta(\tilde \epsilon_1),\delta(\tilde \epsilon_2)]\tilde  \zeta^A =
\ft{1}{2} \bar{\tilde \epsilon}_2\Gamma^a \tilde \epsilon_{1}
\partial_a \tilde{\zeta^A} - \ft{1}{4} \Gamma_a \tilde \Gamma^A \bar{\tilde \epsilon}_2 \Gamma^a
\tilde \epsilon_1.
\end{equation}
The function $\tilde \Gamma^A$ we introduced is given by
\begin{equation} \label{bewvgl6}
\tilde \Gamma^A = \Gamma^a \mathcal{D}_a \tilde \zeta^A + \ft{1}{2}
W_{CDB}{}^A \tilde \zeta^B \bar{\tilde \zeta}^D \tilde \zeta^C.
\end{equation}
This function $\tilde \Gamma^A$ prevents the algebra from closing.
Therefore we again obtain the equations of motion $\tilde{\Gamma}^A = 0$.

\section{From vector multiplet to hypermultiplet} \label{secvechyper}

As explained in the introduction, we will now consider how rigid K{\"a}hler
manifolds in 4-dimensional theories give rise to hyper-K{\"a}hler manifolds
upon dimensional reduction.

We start from the four-dimensional vector multiplet. The geometry
associated with the scalars in this model is known as special geometry.
Rigid special geometry was first introduced in
\cite{Sierra:1983cc,Gates:1984py}. Its ingredients are summarized in
subsection~\ref{ss:N2d4vector}. In subsection \ref{ss:d4tod3} the
dimensional reduction of this model to $3$ dimensions, giving rise to a
hypermultiplet~\cite{DeJaegher:1998ka}  with the associated hyper-K{\"a}hler
geometry, is reviewed. In subsection~\ref{ss:Backd5} we obtain this
geometry in the language that we used before in 5 (or in 6) dimensions.
This is the most symmetric way to describe hypercomplex geometry. We can
thus obtain curvature tensors and $W$-tensors for special hyper-K{\"a}hler
manifolds. The geometry of the vector multiplet is characterized by a
holomorphic function $F(X)$. In this way, we calculate curvature tensors
of special hyper-K{\"a}hler manifolds in terms of this function.

\subsection{The $N=2$, $d=4$ vector multiplet}
\label{ss:N2d4vector}

The four-dimensional vector multiplet consists of a complex scalar $X$, a
vector $A_\mu$ and a fermion field $\Omega_i$. The fermion field carries
a chiral SU($2$) index $i=1,2$. We take the convention that spinors with
a lower index have positive chirality, while spinors with an upper index
have negative chirality. The supersymmetric Lagrangian for $n$ vector
multiplets can be written in terms of a holomorphic function $F(X)$. The
arguments $X^I$ ($I = 1, \cdots, n$) refer to the complex scalar fields
of the $n$ vector multiplets. We restrict our attention to Abelian vector
multiplets, as the non-Abelian part does
not modify the geometry. 
The Lagrangian is a trivial truncation from that written for
superconformal tensor calculus in \cite{deWit:1985px}:
\begin{eqnarray}
{\cal L}_F&=& \rmi \partial _\mu F_I \partial ^\mu  \bar X^I +\ft 14\rmi
F_{IJ}{\cal F}_{\mu\nu}^{-I} {\cal F}^{-J\,\mu\nu}+ \rmi F_{IJ}\bar
\Omega_i^I\slashed{D}
\Omega^{iJ}\nonumber\\
&&- \ft 18\rmi F_{IJ} Y_{ij}^I Y^{ij\,J}
+\ft 14\rmi F_{IJK} Y^{ij\, I}\bar \Omega_i^J\Omega_j^K\nonumber\\
&&-\ft 18\rmi F_{IJK} \varepsilon^{ij}\bar \Omega_i^I \gamma^{\mu\nu\rho}
{\cal F}_{\mu\nu\rho}^{-J} \Omega_j^K+\ft 1{12}\rmi
F_{IJKL}\varepsilon^{ij}\varepsilon^{k\ell } \bar \Omega_i^I \Omega_\ell
^J   \bar \Omega_j^K\Omega_k^L+\hc \label{Lvectorrigid}
\end{eqnarray}
$F_{I_1 \cdots I_k}$ denotes the $k$th derivative of $F$. The Hermitian
conjugate is taken by complex conjugation on the bosonic quantities,
raising or lowering the $i,j$ indices on the spinor $\Omega $ and
replacing anti-self-dual with self-dual tensors
\begin{equation}
 {\cal F}_{\mu\nu}^{\pm I} =\ft12\left( {\cal F}_{\mu\nu}^I\mp \ft12\rmi\varepsilon
 _{\mu\nu\rho\sigma}{\cal F}^{\rho\sigma\,I}\right) ,\qquad {\cal F}_{\mu\nu}^I=2\partial _{[\mu}
 A_{\nu]}^I.
 \label{Fdual}
\end{equation}
The complex scalars $X^I$ parametrize an $n$-dimensional target space
with metric
\begin{equation}
  g_{I \bar{J}} =N_{IJ}\equiv  -\rmi F_{IJ} + \rmi \bar{F}_{IJ},\qquad
N^{IJ} \equiv [N^{-1}]^{IJ}.
 \label{metricN}
\end{equation}
This is a K{\"a}hler space as one can derive the metric from a K{\"a}hler
potential
\begin{equation}
 g_{I\bar{J}} = \frac{\partial^2 K(X,\bar{X})}{\partial X^I \partial \bar{X}^J}
 , \qquad
K(X,\bar{X}) = \rmi X^I \bar{F}_I (\bar{X}) - \rmi \bar{X}^I F_I(X).
\label{kahlerpot}
\end{equation}
This geometry is known as \textit{rigid special K{\"a}hler geometry}.

The supersymmetry transformations for the vector multiplet are
\begin{eqnarray} \label{susytransfvec}
\delta X^I & = & \bar{{\tilde\epsilon}}{}^i \Omega_i^I, \nonumber \\
\delta A^I_\mu & = & \varepsilon^{ij} \bar{{\tilde\epsilon}}_i \gamma_\mu
\Omega^I_j + \varepsilon_{ij} \bar{{\tilde\epsilon}}{}^i \gamma_\mu
\Omega^{jI}, \nonumber \\
\delta \Omega_i^I + \Gamma^I_{JK} \delta X^J \Omega^K_i & = & 2
\slashed{\partial} X^I {\tilde\epsilon}_i - \ft12\rmi \varepsilon_{ij}
\gamma^{\mu \nu} {\tilde\epsilon}^j N^{IJ} \mathcal{G}^{-}_{\mu \nu J} +
\ft{1}{2} \rmi N^{IJ} \bar{F}_{JKL} \bar{\Omega}^{kK} \Omega^{\ell L}
\varepsilon_{ik} \varepsilon_{j\ell} {\tilde\epsilon}^j,
\end{eqnarray}
with $\mathcal{G}_{\mu \nu I}^{-}$ an anti-self-dual tensor defined as
\begin{equation}
\mathcal{G}^{-}_{\mu \nu I} = \rmi N_{IJ} {\cal F}^{-J}_{\mu \nu} -
\ft{1}{4}F_{IJK} \bar{\Omega}^J_i \sigma_{\mu \nu} \Omega^K_j
\varepsilon^{ij}.
\end{equation}
As in section~\ref{ss:hyperd4}, supersymmetry parameters
${\tilde\epsilon}^i$ and ${\tilde\epsilon}_i$ have respectively positive
and negative chirality.

\subsection{Reduction to $d=3$}
\label{ss:d4tod3}

In this section, we pursue by reviewing the reduction to $3$ spacetime
dimensions. A complete discussion can be found in
\cite{DeJaegher:1998ka}. In reducing, the complex scalar $X^I$ loses its
dependence from the fourth spacetime
coordinate. 
An extra scalar $A^I$ from the $\mu=3$ component of the vector appears in
the model. One also obtains a three-dimensional vector. However, in $3$
dimensions vectors are dual to scalars. This scalar can be brought into
the model by use of a Lagrange multiplier. One adds a Lagrange multiplier
term $B_I \varepsilon^{\mu \nu \rho} \partial_\mu F^I_{\nu \rho}$ to the
Lagrangian, in order to impose the Bianchi identity. Integrating the
field strengths, one has introduced the extra scalar $B_I$ in the
Lagrangian. In the case of $n$ vector multiplets, we thus obtain $4n$
real scalars in the reduction process. The resulting geometry is that of
a hyper-K{\"a}hler manifold, i.e. the restriction of the hypercomplex
manifolds considered in section~\ref{sechyper} to manifolds with a
suitable metric. As we work here with a Lagrangian, its kinetic terms of
the scalars guarantee that such a metric exists.

Reducing the spinors gets somewhat more complicated. In four dimensions
the spinors are Majorana spinors with four components. In the reduction
process, each spinor will split up into $2$ spinors with $2$ components.
However we would like to keep on working with spinors with four
components. We will take the following representation of the
$3$-dimensional Clifford algebra
\begin{equation}
\gamma^\mu = \gamma^\mu_{(4)} \tilde{\gamma}, \qquad (\mu = 0,1,2),
\end{equation}
in which $\gamma^\mu_{(4)}$ are the four-dimensional $\gamma$-matrices.
The matrices $\gamma^\mu$ are an alternative set of $4\times 4$ matrices
satisfying the three-dimensional Clifford algebra and commuting with the
remaining $\gamma ^3$. The matrix $\tilde{\gamma}$ contains the unused
four-dimensional $\gamma$-matrices and is defined as
\begin{equation}
\tilde{\gamma} \equiv - \rmi \gamma^3 \gamma^5,
\end{equation}
where $\gamma^3 = \gamma^3_{(4)}, \gamma^5 = \gamma^5_{(4)}$. The
three-dimensional charge conjugation matrix $\mathcal{C}_3$ is given by
\begin{equation}
\mathcal{C}_3 = \mathcal{C}_4 \tilde{\gamma}.
\end{equation}
The reduced Lagrangian can be found in \cite{DeJaegher:1998ka}. Keeping
only the kinetic terms of the scalars we obtain
\begin{eqnarray} \label{geredlagrang}
\mathcal{L} & = & \rmi(\partial_\mu F_I \partial^\mu \bar{X}^I  -
\partial_\mu \bar{F}^I \partial^\mu X^I) \nonumber \\
& & - N^{IJ} (\partial_\mu B_I - F_{IK} \partial_\mu A^K)
(\partial^\mu B_J - \bar{F}_{JM}
\partial^\mu A^M)  + \cdots .
\end{eqnarray}
One can also reduce the supersymmetry transformations. This leads to
\begin{eqnarray} \label{geredsusytransf}
\delta X^I & = & -\rmi \bar{{\tilde\epsilon}}^i \gamma_3 \Omega^I_i, \nonumber\\
\delta A^I & = & \rmi \varepsilon^{ij} \bar{{\tilde\epsilon}}_i
\Omega^I_j - \rmi
\varepsilon_{ij} \bar{{\tilde\epsilon}}^i \Omega^{jI}, \nonumber \\
\delta B_I & = & \rmi F_{IJ} \varepsilon^{ij} \bar{{\tilde\epsilon}}_i
\Omega^J_j - \rmi \bar{F}_{IJ} \varepsilon_{ij} \bar{{\tilde\epsilon}}^i
\Omega^{jJ}, \nonumber \\
\delta \Omega^I_i & = & 2 \rmi \gamma^\mu \partial_\mu X^I \gamma_3
{\tilde\epsilon}_i + 2 N^{IJ}(\gamma^\mu \partial_\mu B_J - \bar{F}_{JK}
\gamma^\mu \partial_\mu A^K) \varepsilon_{ij} {\tilde\epsilon}^j
\nonumber
\\
& & + \rmi N^{IJ} \delta F_{JK} \Omega^K_i - N^{IJ} \bar{F}_{JKL}
N^{KM} (\delta B_M - F_{MN} \delta A^N) \varepsilon_{ij} \gamma_3
\Omega^{Lj}, \nonumber \\
\delta \Omega^{Ii} & = & - 2 \rmi \gamma^\mu \partial_\mu \bar{X}^I
\gamma_3 {\tilde\epsilon}^i + 2 N^{IJ}(\gamma^\mu \partial_\mu B_J -
F_{JK} \gamma^\mu \partial_\mu A^K) \varepsilon^{ij} {\tilde\epsilon}_j
\nonumber
\\
& & - \rmi N^{IJ} \delta \bar{F}_{JK} \Omega^{Ki} - N^{IJ} F_{JKL} N^{KM}
(\delta B_M - \bar{F}_{MN} \delta A^N) \varepsilon^{ij} \gamma_3
\Omega^L_j.
\end{eqnarray}
Note that we keep the fermion fields in their original four-dimensional
form. They are doublets of $\frac{1}{2}(1 \pm \gamma_5)$ projections of
four-dimensional Majorana spinors. However, the definition of the
conjugate of a spinor has been modified according to the explained rules.

\subsection{Back to $d=5$}
\label{ss:Backd5}

In this section, we will lift the supersymmetry transformation rules for
the $3$-dimensional hypermultiplet up to $5$ dimensions. In this way, we
are able to write down the corresponding geometrical quantities,
associated with the hyper-K{\"a}hler geometry, such as the vielbeins and the
spin connections in this specific case.

We use the following identification for the spinors in order to find
transformation rules that are consistent with those of the
five-dimensional hypermultiplet:
\begin{equation} \label{idzetaomega}
\zeta^{1I} = \gamma_3 \Omega^{1I} + \Omega^{2I}, \qquad \zeta^{2I} =
\Omega^I_2 - \gamma_3 \Omega^I_1,
\end{equation}
where we used $\zeta$ to denote the five-dimensional spinors. For the
supersymmetry parameters, we can give an analogous identification
\begin{equation} \label{ideps2}
\epsilon^1 = \gamma_3 \tilde \epsilon_1 + \tilde \epsilon_2, \qquad
\epsilon^2 = \tilde \epsilon^2 - \gamma_3 \tilde \epsilon^1,
\end{equation}
where we used $\epsilon$ to denote the five-dimensional parameters. Using
the fact that $\mathcal{C}_3 = \mathcal{C}_4 \tilde{\gamma} = \rmi
\mathcal{C}_5 \gamma_3$, we can obtain the following transformation rules
for the scalars:
\begin{eqnarray} \label{geredsusytransf5}
\delta x^I & = & \frac{1}{2}\bar{\epsilon}^2 \zeta^{2I} -
\frac{1}{2} \bar{\epsilon}^1 \zeta^{1I}, \nonumber \\
\delta y^I & = & - \frac{\rmi}{2}\bar{\epsilon}^2 \zeta^{2I} -
\frac{\rmi}{2} \bar{\epsilon}^1 \zeta^{1I}, \nonumber \\
\delta A^I & = & - \bar{\epsilon}^1 \zeta^{2I} -
\bar{\epsilon}^2 \zeta^{1I}, \nonumber \\
\delta B_I & = & - F_{IJ} \bar{\epsilon}^1 \zeta^{2J} - \bar{F}_{IJ}
\bar{\epsilon}^2 \zeta^{1J},
\end{eqnarray}
where $x^I = \Re X^I$ and $y^I = \Im X^I$.

\subsection{Some geometrical quantities}

Now that we have reduced the supersymmetry transformations to five
dimensions, we can compare these rules with the general transformation
rules for a five-dimensional hypermultiplet. In this way, we can extract
the necessary quantities that characterize the hyper-K{\"a}hler geometry. We
are especially interested in determining the vielbeins and spin
connections.

We first compare the transformations (\ref{geredsusytransf5}) with the
general transformation rules (\ref{susyD5}). Remember that the general
index $A$ is here represented by a combination $iI$, with $i=1,2$ and
$I=1,\ldots n$. The vielbeins are
\begin{eqnarray}
 f^{1\,1I} & = & \rmi \rmd x^I +\rmd y^I \nonumber\\
 f^{1\,2I} & = & N^{IJ}\left(\rmd B_J-\bar F_{JK}\rmd A^K\right)\nonumber\\
f^{2\,1I} & = & N^{IJ}\left(-\rmd B_J+ F_{JK}\rmd A^K\right)\nonumber\\
 f^{2\,2I} & = &-\rmi \rmd x^I +\rmd y^I.
 \label{vielbeinsc}
\end{eqnarray}
We present the inverse vierbeins, which take a simpler form as a matrix
whose rows represent the components
\begin{equation}
  \pmatrix{x^I \cr y^I\cr A^I\cr B_I}, 
 \label{componentsorderX}
\end{equation}
and the columns the $(iA)$ values $(1\,1J)$, $(1\,2J)$, $(2\,1J)$ and
$(2\,2J)$:
\begin{equation}
  f^X{}_{iA}=
\left( \begin{array}{cccc}
-\frac{\rmi}{2} \delta^I_J&0&0&\frac{\rmi}{2} \delta^I_J\\ [2mm]
  \frac{1}{2} \delta^I_J&0&0&\frac{1}{2} \delta^I_J\\ [2mm]
  0&-\rmi\delta ^I_J& -\rmi\delta ^I_J& 0\\ [2mm]
  0& -\rmi F_{IJ}&-\rmi\bar F_{IJ} & 0\end{array}\right).  
 \label{invvielbeinc}
\end{equation}

%

We can also obtain the spin connection as matrix in the basis
$A=(1I),(2I)$ (for the rows), $B=(1J),(2J)$ (for the columns)
\begin{equation}
  \omega _{XA}{}^B \rmd q^X=
  \left( \begin{array}{cc}
    N^{JK}\bar F_{IKM} \left(\rmi \rmd x^M+\rmd y^M\right) & N^{JK}\bar F_{IKL} N^{LM}
    \left(-\rmd B_M+ F_{MN}\rmd A^N\right)  \\
    N^{JK}F_{IKL} N^{LM}
    \left(\rmd B_M-\bar F_{MN}\rmd A^N\right)  & N^{JK}F_{IKM}\left(-\rmi \rmd x^M+\rmd y^M\right)
  \end{array}\right) .
 \label{spinconnectionc}
\end{equation}

%
%
%
We now have all the necessary information. The metric can be read off
from the Lagrangian~(\ref{geredlagrang})
\begin{eqnarray}
  g_{XY}\rmd q^X\rmd q^Y&=& N_{IJ}\rmd x^I\rmd x^J+N_{IJ}\rmd y^I\rmd
  y^J+\ft14N_{IJ}\rmd A^I\rmd
  A^J \nonumber\\
  && +N^{IJ}\left( \rmd B_I-\Re F_{IK}\rmd A^K\right)\left( \rmd B_J-\Re F_{JL}\rmd
  A^L\right).
 \label{metricc}
\end{eqnarray}
Using this metric, one can calculate the Levi-Civita connection. Since we
are dealing with a hyper-K{\"a}hler manifold, this connection coincides with
the Obata connection (i.e. the connection defined on hypercomplex
manifolds that leaves the complex structures invariant). The complex
structures can also be calculated. In a basis where the rows and the
columns are represented by (\ref{componentsorderX}), they are:
\begin{eqnarray} \label{complstructsphyp}
J^1 & = & \left( \begin{array}{cccc} 0 & 0 & 0 & -N_{IJ} \\ 0 & 0 & -2
\delta_I^J & -2 \Re F_{IJ} \\ -N^{JL} \Re F_{LI} & \frac{1}{2} \delta_I^J
& 0 & 0 \\ N^{IJ} & 0 & 0 & 0 \end{array} \right),\\ J^2 & = & \left(
\begin{array}{cccc} 0 & 0 & -2\delta_I^J & -2\Re F_{IJ} \\ 0 & 0 & 0 &
N_{IJ} \\ \frac{1}{2} \delta_I^J & N^{JL} \Re F_{LI} & 0 & 0 \\ 0 &
-N^{IJ} & 0 & 0
\end{array} \right),\\
J^3 & = & \left( \begin{array}{cccc} 0 & \delta_I^J & 0 & 0 \\
-\delta_I^J & 0 & 0 & 0 \\ 0 & 0 & 2N^{JL} \Re F_{LI} & 2\Re( F_{IL}
N^{LK} \bar F_{KJ})  \\ 0 & 0 & -2N^{IJ} & -2N^{IK} \Re F_{JK}
\end{array} \right).
\end{eqnarray}

Since we are working with a hyper-K{\"a}hler manifold, we can also define a
fermion metric
\begin{equation}
C_{CD} = \ft{1}{2}\varepsilon^{ij} f^X_{iC} g_{XY} f^Y_{jD},
\end{equation}
which results here in
\begin{equation}
C_{iI \, jJ} = \varepsilon _{ij}N_{IJ} . \label{Cmetric}
\end{equation}

\subsection{The $W$-tensor}

The tensor $W_{ABCD}$ with the last index lowered using  the metric
(\ref{Cmetric}) is symmetric. It can be calculated using the curvature
relations (\ref{defW}). The result can be represented using arbitrary
symplectic vectors $A^A=A^{iI}$:
\begin{eqnarray}
  W_{ABCD}A^AA^BA^CA^D&=& -3V_I
  N^{IJ}V_J-\rmi A^{1I}A^{1J}A^{1K}A^{1L}\bar F_{IJKL}+\rmi A^{2I}A^{2J}A^{2K}A^{2L} F_{IJKL}
  \nonumber\\
 V_I&\equiv &\bar F_{IKL}A^{1K}A^{1L}+F_{IKL}A^{2K}A^{2L}\label{Wc}
\end{eqnarray}
Using relations  (\ref{RGROm}) and (\ref{defW}), it is clear that the
$W$-tensor determines the other curvature tensors using the vielbeins
that were obtained in the previous section. 
Therefore, this formula contains the full description of the geometry of
the special hyper-K{\"a}hler manifolds. Moreover this tensor also contains
the information about the dynamical properties of the system, since it
appears explicitly in the equations of motion for the fermions (see
(\ref{Gamma})). The equations of motion for the scalars can be derived
from those for the fermions by application of a supersymmetry
transformation.

\section{Conclusions}

In the first part of this paper (section \ref{sechyper}), we made the
connection between formulations of hypermultiplets in 4, 5 and 6
dimensions. We showed how the same geometrical relations appear in
different settings and that these are equivalent by giving explicit
translation formulae. Thus, this connects different formulations of the
geometry of hypercomplex manifolds, appropriate for supersymmetry in
different dimensions. We did not assume the existence of an action. This
possibility was already shown in the five-dimensional case
in~\cite{Bergshoeff:2002qk} through closure of the supersymmetry algebra,
and is hereby extended to $4$ and $6$ dimensions. In every case, we were
able to link the vielbeins, appearing in the supersymmetry transformation
rules, with the five-dimensional ones. In each dimension we have also
obtained equations of motion through closure of the supersymmetry
algebra.

In the second part of this paper (section \ref{secvechyper}), we have
further investigated the dimensional reduction of the rigid
four-dimensional vector multiplet to 3 dimensions as performed
in~\cite{DeJaegher:1998ka}. Connecting also the 3-dimensional to the
5-dimensional formulation, simplifies identifications of geometric
quantities. As the special geometry of 4 dimensions is fixed by the
holomorphic function $F(X)$ we obtain the curvatures and invariant
tensors of a subclass of hyper-K{\"a}hler manifolds also in terms of this
function. This subclass are the `special hyper-K{\"a}hler' manifolds. We
obtain all results in a uniform language such that the general analysis
of properties of hypercomplex manifolds that was made in appendix~B
of~\cite{Bergshoeff:2002qk} can be applied. In particular we obtain the
$W$-tensor of special hyper-K{\"a}hler manifolds.

\medskip
\section*{Acknowledgments}

\noindent

We are grateful to Jos Gheerardyn and Stefan Vandoren for interesting
discussions. This work is supported in part by the European Community's
Human Potential Programme under contract HPRN-CT-2000-00131 Quantum
Spacetime. J.R. is Aspirant FWO-Vlaanderen. The work is supported in part
by the Federal Office for Scientific, Technical and Cultural Affairs
through the `Interuniversity Attraction Poles Programme -- Belgian
Science Policy' P5/27.

\providecommand{\href}[2]{#2}\begingroup\raggedright\endgroup

\end{document}